\documentclass[superscriptaddress]{revtex4-1}

\usepackage[english]{babel}
\usepackage{graphicx}
\usepackage{amsmath}
\usepackage{textcomp}

\usepackage{xcolor}     
\usepackage{soul}       

\begin{document}

\begin{huge}
\noindent{\textbf{Velocity-dependent quantum phase slips in 1D atomic superfluids\\}}
\end{huge}

\begin{large}
\noindent{\textbf{Luca Tanzi$^{1,+}$, Simona Scaffidi Abbate$^{1,+}$, Federica Cataldini$^1$, Lorenzo Gori$^1$, Eleonora Lucioni$^1$, Massimo Inguscio$^{1,2}$, Giovanni Modugno$^{1,2}$ \& Chiara D'Errico$^{1,2,*}$\\}}
\end{large}

\begin{small}
\noindent{$^1$ LENS and Dipartimento di Fisica e Astronomia, Universit\'a di Firenze, 50019 Sesto Fiorentino, Italy}\\
\noindent{$^2$ Istituto Nazionale di Ottica, CNR, 50019 Sesto Fiorentino, Italy}\\
\noindent{$^*$ derrico@lens.unifi.it}\\
\noindent{$^+$these authors contributed equally to this work}\\
\end{small}

\noindent{\textbf{Quantum phase slips are the primary excitations in one-dimensional superfluids and superconductors at low temperatures but their existence in ultracold quantum gases has not been demonstrated yet. We now study experimentally the nucleation rate of phase slips in one-dimensional superfluids realized with ultracold quantum gases, flowing along a periodic potential. We observe a crossover between a regime of temperature-dependent dissipation at small velocity and interaction and a second regime of velocity-dependent dissipation at larger velocity and interaction. This behavior is  consistent with the predicted crossover from thermally-assisted quantum phase slips to purely quantum phase slips.\\}}


\noindent{Phase slips, i.e. phase fluctuations of the superfluid order parameter, are the dominant excitations of one-dimensional (1D) superfluids and superconductors in the presence of an obstacle for the superflow.} Remarkably, phase slips may occur even at zero temperature, due to quantum tunneling events \cite{Giordano}. This mechanism, known as quantum phase slips (QPS), controls the dissipation of nominally frictionless systems and has been observed in quasi-1D superconducting nanowires \cite{Bezryadin01,Lau01,Altomare,Bezryadin09,Kamenev,BezryadinRev} and in Josephson junctions chains \cite{Pop10}. There is currently a large interest in QPS as the fundamental process for the realization of topologically protected qubits \cite{Mooij05, Astafiev} or for the implementation of a quantum standard for the electrical current \cite{Mooij06,Pop10}.
QPS have not yet been clearly identified in superfluids based on ultracold quantum gases. Although several studies have shown the presence of strong dissipation for superfluids moving in optical lattices \cite{Fertig05,Ketterle07,Demarco08,Tanzi}, the closest indication of QPS is just the onset of a regime of temperature-independent dissipation at low temperature \cite{Demarco08}. In this work we study for the first time how the dissipation depends on the superfluid velocity, in addition to other key parameters such as temperature and strength of the interparticle interaction. We observe a clear crossover between a temperature-dependent regime and a velocity dependent regime, in general agreement with theoretical predictions for the crossover from thermal to quantum phase slips \cite{Buchler01,Polkovnikov05,Polkovnikov12,Danshita13}. This indicates that QPS can be observed and controlled also in ultracold quantum gases.

\begin{figure}[ht!]
\centering{\includegraphics[width=0.65\columnwidth] {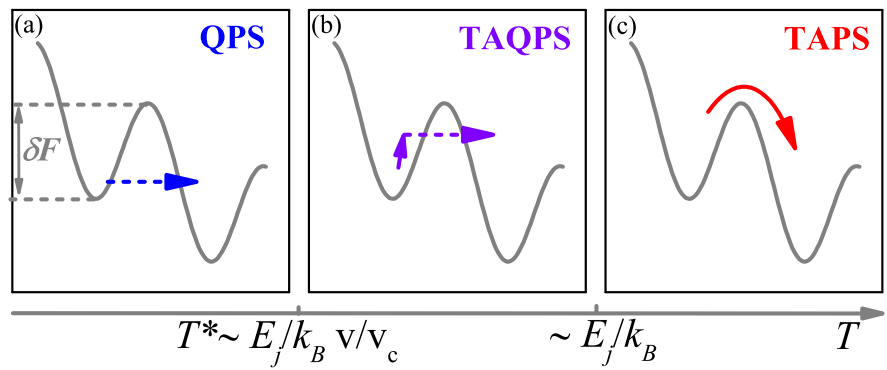}}
\caption{\textbf{Scheme of the different phase-slips activating mechanisms as a function of temperature.} Phase-slips are activated (a) via quantum tunneling events (QPS) for $T\lesssim T^*$, (b) via quantum tunnelling events assisted by the temperature (TAQPS) for  $T^* \lesssim T \lesssim\delta F/k_B$ and (c) via thermal fluctuations (TAPS) for $T\gg\delta F/k_B$. In a lattice $T^*\simeq E_j/k_B \times v/v_c$ and $\delta F \simeq E_j$, see text.}
\label{fig1}
\end{figure}

Let us start by introducing the mechanisms for the generation of phase slips. A 1D superfluid can be described by a complex order parameter $\Psi(x)=|\Psi(x)|e^{i\phi(x)}$. The superfluid metastable state corresponds to a local minimum of the Ginzburg-Landau free energy potential $F$ \cite{Landau}. A phase slip event is a local fluctuation in $\Psi(x)$ corresponding to the suppression of its modulus and a simultaneous phase jump of 2$\pi$. When a phase slip occurs, the state with superfluid velocity $v\propto\nabla\phi(x)$ decays into a state with lower velocity, since the phase has locally unwound \cite{Little}. As shown in Fig.~\ref{fig1}, three different processes may activate a phase slip, depending on the temperature regime. When the temperature is much higher than the free-energy barrier between two metastable states, $T\gg\delta F/k_B$, the order parameter may overcome the barrier via thermal fluctuations, causing the formation of thermally activated phase slips (TAPS) with a nucleation rate following the Arrhenius law $\Gamma\propto e^{-\delta F/k_BT}$ \cite{Langer,McCumber}. When $T\lesssim\delta F/k_B$, the probability of TAPS becomes small, and phase slips occur mainly via quantum tunnelling through the free-energy barrier. Following quantum mechanical arguments one can find a characteristic temperature $T^*$ below which the QPS nucleation rate is temperature-independent \cite{Giordano,Arutyunov}, while in an intermediate temperature range, $T^*\lesssim T\lesssim\delta F/k_B$, QPS are thermally assisted (TAQPS) \cite{Weiss}.

The analytical form of $\delta F$ and $T^*$ depends on the specific type of obstacle experienced by the superflow, e.g. disorder, isolated defects or periodic potentials \cite{Pryadko,Buchler01,Polkovnikov12}. For a superfluid moving along a periodic potential the relevant energy scale is the Josephson plasma energy $E_j$ \cite{Danshita13,Polkovnikov12}, which sets the free-energy barrier $\delta F \simeq E_j$ and determines also the crossover temperature, $T^*\simeq E_j/k_B \times v/v_c$, between the QPS and TAQPS regimes. Here $v$ and $v_c$ are the superfluid velocity and the critical velocity for the dynamical instability \cite{Smerzi,WuNiu,Fallani}, respectively. Theoretical studies in the Bose-Hubbard limit show that the phase-slip nucleation rate should have a characteristic dependence on velocity, temperature and interaction strength that is quite different for TAQPS and QPS: $\Gamma\sim vT^{2K-3}$ for the TAQPS regime and $\Gamma\sim v^{2K-2}$ for the QPS regime, where the Luttinger parameter $K$ is related to the interaction strength \cite{Danshita13,BuchlerT,Polkovnikov12}. These results apply for small velocities, far from the critical velocity $v_c$, where instead TAPS tend to dominate \cite{Polkovnikov05}.

Experiments with ultracold gases in optical lattices have revealed a strong dissipation that may be associated to phase slip proliferation \cite{Fertig05,Ketterle07,Demarco08,Tanzi}. Most of the experiments have been performed for large velocity or large interaction, close to the dynamical instability \cite{Ketterle07,Tanzi} or to the superfluid-insulator transition \cite{Fertig05,Ketterle07,Tanzi}. Characterizing the phase slips is challenging in such limit, since the nucleation rates tend to diverge \cite{Polkovnikov05}. At lower velocities, a weaker dissipation that depends on the interaction has been observed \cite{Demarco08,Tanzi}, as well as the onset of a regime of temperature-independent dissipation \cite{Demarco08}. The latter in particular is a strong indication of the onset of QPS, although it has not been possible to find a quantitative agreement with the theory. So far, it has not been possible to study the velocity dependence of the dissipation rate predicted for the TAQPS and QPS regimes.

In this work we demonstrate that it is possible to measure also the velocity dependence, by employing 1D superfluids with tunable interaction and weak periodic potentials. The idea is that a weak lattice in the Sine-Gordon limit shifts the critical velocity towards the band edge, thus enlarging the range of accessible $v$. The tunable interaction adds instead an independent way of controlling the Josephson energy. By tuning both velocity and interaction we indeed observe a clear change of behavior of the phase-slip nucleation rate that resembles that of the theory and suggests a crossover between the TAQPS and QPS regimes.\\

\begin{large}
\noindent{\textbf{Experimental observables}}
\end{large}

\noindent{In the experiment we employ an array of 1D superfluids in an optical lattice. In order to study the dissipation rate, we excite oscillations of the superfluids by suddenly displacing the center of an harmonic trap that confines the atoms, as shown in Fig.~\ref{fig2}a.
\begin{figure}[h!]
\centering
   {\includegraphics[width=1\columnwidth] {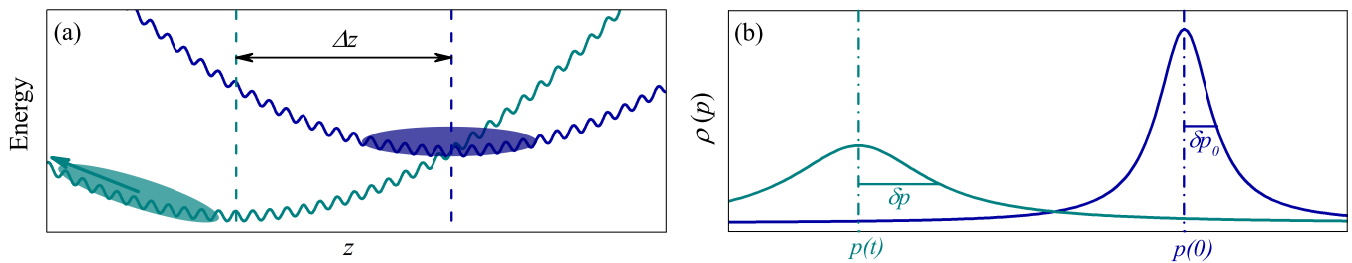}}
\caption{\textbf{Scheme of the experimental sequence.} Not to scale. (a) By displacing the harmonic trap center at $t=0$ (blue), we excite an oscillation of the 1D system in the shifted potential (cyan). $\Delta z$ is the diplacement between the equilibrium positions (dashed lines) of the two potentials. (b) Cartoon of momentum distributions at $t=0$ (blue) and at a variable evolution time (cyan). $p$ is measured as the shift of the distributions' centers (dash-dotted lines). $\delta p_0$ and $\delta p$ are the momentum distribution widths at $t=0$ and $t \neq 0$, respectively.}
\label{fig2}
\end{figure}
By changing the displacement $\Delta z$ we can excite oscillations with different amplitudes. After a variable oscillation time, we suddenly switch off all the confining potentials and we let the atoms free to expand to record the momentum distribution $\rho(p)$ (Fig.~\ref{fig2}b). Since the expansion mixes all subsystems, our measurements give information on their mean properties. By fitting $\rho(p)$ with a Lorentzian function, we measure the quasi-momentum $p$ and the half-width at half maximum, $\delta p$. Since the lattice dispersion is linear in the whole velocity range we have studied, we can safely identify $p$ with the center-of-mass momentum. From the momentum width at $t$=0, $\delta p_0$, we can estimate the temperature \cite{Gerbier,Gerbier04}, which in our measurements ranges from 20 to 40~nK. The measurements are performed for a wide range of interaction strength; in terms of the Lieb-Liniger parameter $\gamma$, this ranges from 0.13 to 1.22. The Josephson plasma energy $E_j$ depends on the interaction strength via the sound velocity [see Methods]. Moving from weak to strong interactions, $E_j/k_b$ varies from 20 to 35 nK.
\begin{figure}[th!]
\centering
   {\includegraphics[width=0.6\columnwidth] {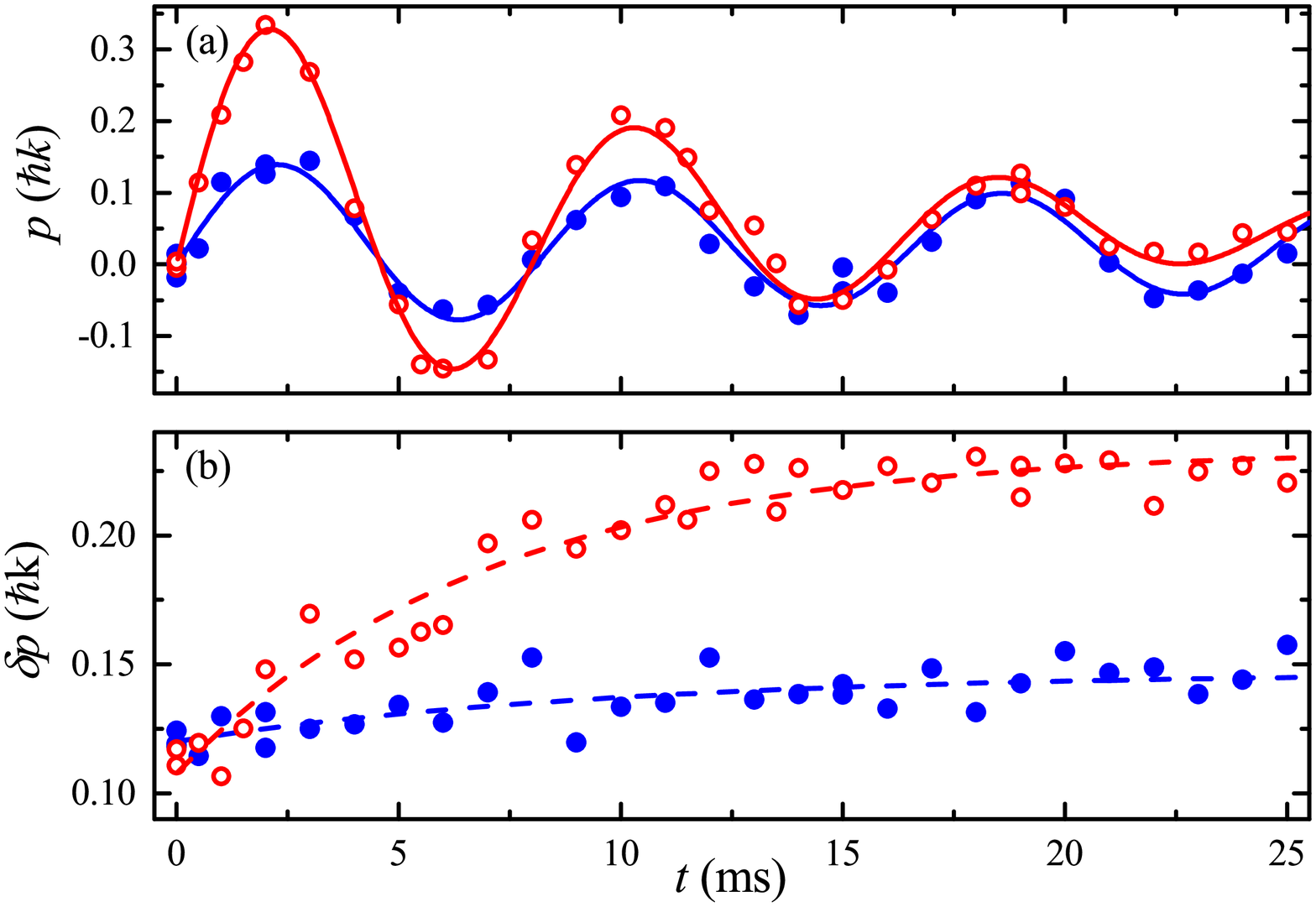}}
\caption{\textbf{Damped oscillation of the array of 1D superfluids in the optical lattice.} (a) Time evolution of the quasi-momentum $p$ and (b) of the momentum distribution width $\delta p$  for the interaction strength $\gamma =1.22$, the temperature $T= 22(4)$~nK and two maximum velocities: $v=1.4(4)$~mm/s (blue filled circles) and $v=2.2(4)$~mm/s (red open circles), respectively corresponding to trap displacements $\Delta z$=1.5~$\mu$m and $\Delta z$=4~$\mu$m. The lines in panel (a) are fits to measure the damping rate, which is $G=28(9)$~Hz and $G=84(6)$~Hz for the blue and red data, respectively. The lines in panel (b) are fits to measure the time constant $\tau$, which is $\tau=10(7)$~ms and $\tau=7(1)$~ms for the blue and red data, respectively.}
\label{fig3}
\end{figure}
Fig.~\ref{fig3} shows a typical observation for the time evolution of $p$ and $\delta p$ at a given interaction ($\gamma = 1.22$) and at two different trap displacements, which correspond to two different velocities.  In accordance to the theory, we label each dataset with the maximum velocity $v$ reached during the first oscillation \cite{Danshita13}.
We fit the evolution of $p$ using an oscillator model with a friction term: $p(t)=m^*\tilde{v} \exp(-Gt)\sin(\omega't+\varphi)$, where  $\omega'=\sqrt{\omega_z^2m/m^*-G^2}$, $m^*$ is the effective mass in the lattice, $\omega_z$ is the frequency of the harmonic potential, $\tilde{v}$ and $\varphi$ are fitting parameters [see Methods]. The growth of $\delta p$ is fitted with an inverted exponential with time constant $\tau$. We observe that the damping rate $G$ is directly related to $\tau$, via $2\tau\simeq1/G$ (within 0.6 standard deviations), as the mechanical energy dissipated in the oscillation is converted into momentum spread.\\

\begin{large}
\noindent{\textbf{Results}}
\end{large}

\noindent{By measuring the time evolution of $p$ we have direct access to $G$, the damping rate of the oscillation. The theory predicts a direct relation between the measured damping rate $G$ and the phase slip nucleation rate $\Gamma$, i.e.~$G\propto \Gamma/v$ \cite{Danshita13,note2}. Therefore, in the QPS regime the damping rate $G$ depends on $v$ but not on $T$, $G\sim v^{2K-3}$. Conversely, in the TAQPS regime $G$ depends on $T$ but not on $v$, $G\sim T^{2K-3}$.

\begin{figure}[h!]
\centering
   {\includegraphics[width=0.6\columnwidth] {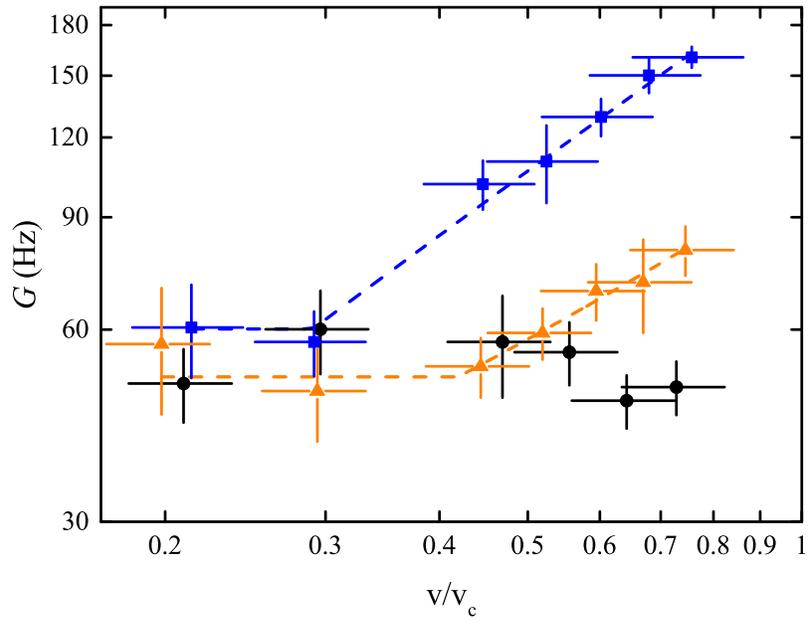}}
\caption{\textbf{Velocity dependence of the damping rate for various interaction strengths}. The damping rate $G$ is plotted vs the maximum velocity $v$ normalized to the critical velocity $v_c$, for three interaction strengths and constant temperature: $\gamma = 0.13$ and $T = 37(7)$ nK (black circles), $\gamma = 0.19$ and $T = 39(7)$ nK (orange triangles) and $\gamma = 0.64$ and $T = 34(5)$ nK (blue squares). The lines are piece-wise linear fits to determine the crossover velocity $v^*$ (see text). The error bars represent the statistical uncertainties.}
\label{fig4}
\end{figure}

As shown in the example in Fig.~\ref{fig3} for $\gamma = 1.22$, a small velocity $v$ typically leads to a very weakly damped oscillation. A larger $v$ with the same interaction strength leads instead to a larger damping, suggesting the presence of phase slips with a nucleation rate that depends on the velocity. We have repeated this type of measurement for a wide range of velocities and interaction strengths. A summary of the evolution of $G$ with velocity and interaction for approximately constant temperature is shown in Fig.~\ref{fig4}. Each data set corresponds to a different $\gamma$ and is rescaled to the corresponding critical velocity $v_c$ for the occurrence of the dynamical instability. At weak interaction (black), $G$ is essentially independent of $v$.  At stronger interactions (orange, blue), we observe instead a clear crossover from a regime of constant $G$ to a regime where $G$ grows with the velocity. A fit of the data with a piece-wise linear function is used to determine the crossover velocity $v^*$, that is the minimum velocity required to enter the regime of dependence on $v$. The crossover velocity apparently decreases for increasing interaction.

\begin{figure}[h!]
\centering
   {\includegraphics[width=0.6\columnwidth] {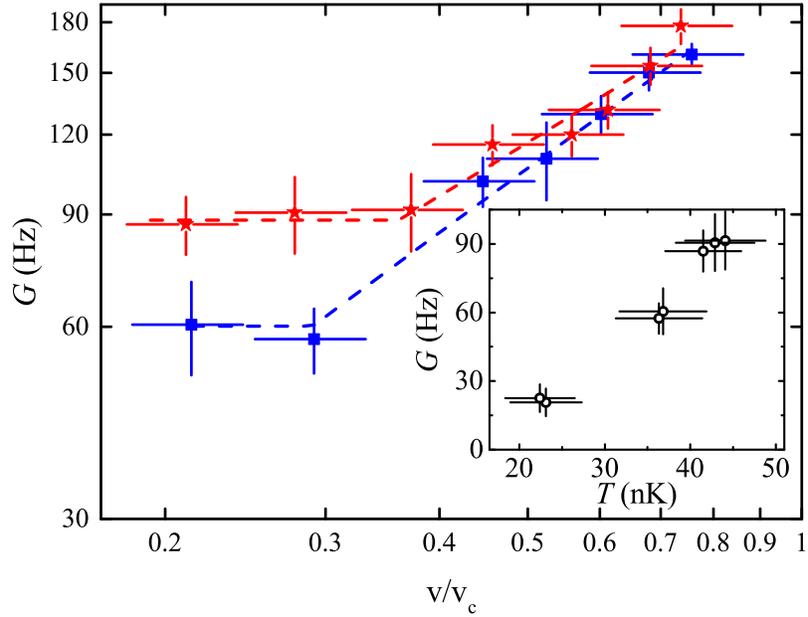}}
\caption{\textbf{Effect of temperature on the damping rate.} $G$ is plotted vs $v/v_c$ for two different temperatures and approximately constant interaction energy: $T = 34(5)$ nK and $\gamma = 0.64$ (blue squares) and $T = 43(5)$ nK and $\gamma = 0.70$ (red stars). The lines are piece-wise linear fits  to determine the crossover velocity $v^*$ (see text). Inset: Temperature dependence of the damping rate at small velocities ($v<v^*$). The error bars represent the statistical uncertainties.}
\label{fig5}
\end{figure}
A similar behavior is observed also at different temperatures. For example in Fig.~\ref{fig5} we compare two datasets with approximately constant $\gamma$ but different $T$. In the $v$-independent regime the damping rate $G$ is strongly affected by temperature, with a monotonic increase of $G$ with $T$ (inset in Fig.~\ref{fig5}), while the dependence on interaction (Fig.~\ref{fig4}) is weaker. Instead, in the $v$-dependent regime interaction effects are apparently dominant (Fig.~\ref{fig4}) and we cannot measure a clear dependence on $T$ (Fig.~\ref{fig5}). The combination of these effects leads to an increase of the crossover velocity for increasing temperature.\\

\begin{large}
\noindent{\textbf{Discussion}}
\end{large}

These observations suggest that our system is at the crossover between the TAQPS and QPS regimes. As we will show, the crossover is not controlled by changing the temperature $T$, but rather by varying the crossover temperature $T^*\propto E_j /k_B \times v/v_c$ through a change in velocity and interaction strength \cite{Polkovnikov12}. For $T^*<T$, i.e. at small velocity and small interaction, the system is apparently in the TAQPS regime, since $G$ does not show any substantial dependence on $v$, but only on $T$. For $T^*>T$, i.e. at large velocity and large interaction, the system enters a regime where $G$ is temperature independent and is approximately linearly dependent on the velocity. This suggests that the system is in a regime of QPS. A further indication of the observation of the TAQPS-QPS crossover comes from the dependence of $v^*$ on temperature and interaction strength. In Fig.~\ref{fig6} we report the measured crossover velocity normalized to the critical velocity, $v^*/v_c$, versus the temperature normalized to the Josephson energy, $k_BT/E_j$. The data show a clear linear scaling, which is consistent with the theoretical prediction $k_BT^*/E_j\propto v/v_c$. From a fit we get $k_BT^*$=4.9(14)$E_jv/v_c$ - 0.4(4)$E_j$.

The $v$-dependent regime shows also the characteristics features already observed in previous experiments: a weak $T$ dependence \cite{Demarco08} and a strong interaction dependence \cite{Demarco08,Tanzi}.This suggest that the QPS regime has already been reached in previous experiments on ultracold quantum gases.

\begin{figure}[h!]
\centering
{\includegraphics[width=0.5\columnwidth] {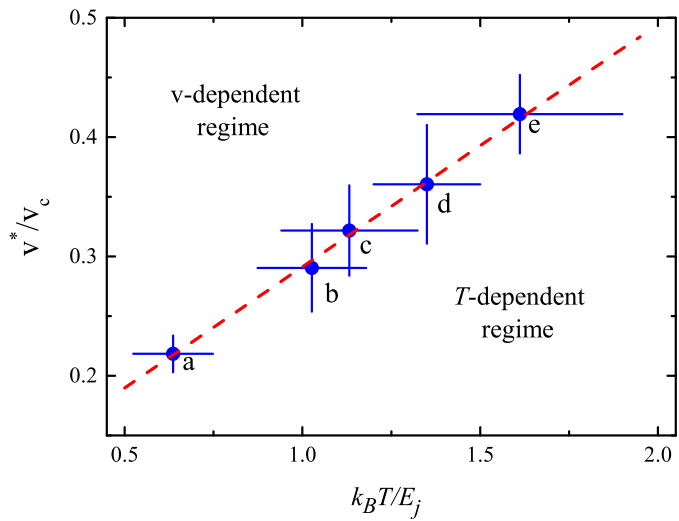}}
\caption{\textbf{Evolution of the crossover velocity in the velocity-temperature plane.} The individual datapoints have been taken for different temperatures and interaction energies: a) $\gamma = 1.22$ and $T = 22(2)$ nK, b) $\gamma = 0.64$ and $T = 34(5)$ nK, c) $\gamma = 0.37$ and $T = 30(5)$ nK, d) $\gamma = 0.70$ and $T = 43(5)$ nK, e) $\gamma = 0.19$ and $T = 39(7)$ nK. The dashed line is the linear fit described in the text, which separates the $v$-dependent and the $T$-dependent regimes. The error bars represent the statistical uncertainties.}
\label{fig6}
\end{figure}

We note that it is not possible to reach a quantitative agreement with the theory for $G(v)$ in the regime we attribute to QPS. On the theory side, the power-law behavior we showed above is valid only for very low velocities ($p<hk/10$), while in the experimental range ($hk/10<p<hk/2$) a different, exponential behavior has been predicted \cite{Danshita13,Polkovnikov12}. Furthermore, the theory is done in the Bose-Hubbard limit, and it is not clear if it can be immediately extended to the Sine-Gordon limit. On the experimental side, the limited range of accessible velocities, which is limited by the finite $T$ on the low-$v$ side, does not allow us to distinguish a power law from an exponential. If we fit our data with a power law, we get exponents of the order unity, which are essentially interaction independent \cite{note5}. Further work in theory and experiment is clearly needed to try reaching a quantitative agreement.

We finally note that our measurements are incompatible with dissipation of the Landau type \cite{Landau41,WuNiu,Fallani,DeSarlo}, which may appear for velocities larger than the sound velocity $v_s$. Indeed, not only we typically have $v< v_s$, but additionally $v_s$ increases with the interaction strength, so that one should observe a decrease of $G$ with increasing $\gamma$ at a fixed velocity, and not the increase shown for example in Fig.~\ref{fig4}.\\

\begin{large}
\noindent{\textbf{Conclusion}}
\end{large}
In conclusion, our measurements reveal a crossover behavior of the dissipation in atomic superfluids that reproduces closely the crossover between TAQPS and QPS predicted by the theory. One important feature of our setup is that the QPS regime can be reached at constant temperature, by tuning the velocity or the interaction. This offers the possibility to control the QPS nucleation rate and opens new perspectives for the study of QPS-related phenomena in ultracold quantum gases. In the future it might be possible to directly observe individual QPS events by interferometric means \cite{Hadzibabic,Schmiedmayer} or by single atom detection \cite{Greiner,Bloch}, allowing to study in depth the nucleation mechanisms of quantum and thermal phase slips.
Furthermore, methods analogous to those reported in this work could be used to study fundamental aspects of the QPS nucleation by individual defects or controlled disorder \cite{Buchler01,Pryadko}.\\

We acknowledge discussions with N. Fabbri, C. Fort and I. Danshita. This work was supported by the ERC (Grant No. 247371 - DISQUA), by the EU - H2020 research and innovation programme (Grant No. 641122 - QUIC) and by the Italian MIUR (Grant No. RBFR12NLNA - ArtiQuS).\\

\begin{large}
\noindent{\textbf{Methods}}
\end{large}

\noindent{In the experiment we employ $^{39}$K atoms, for which we can accurately tune the interaction thanks to a broad Feshbach resonance \cite{Roati07}. The 1D superfluids are realized by splitting a 3D Bose-Einstein condensate into about 1000 subsystems, using a deep 2D lattice in the horizontal plane \cite{Tanzi}.} Each subsystem contains on average 30 atoms. The transverse trapping energy $\hbar\omega_{\perp}$=$h\times$40~kHz is much larger than all other energy scales, realizing effectively one-dimensional systems. The weak optical lattice with depth $V$=$1.0(1) E_R$ is then added along the longitudinal direction $z$. The lattice spacing is $d$=532~nm and $E_R=\hbar^2k^2/2m$ is the recoil energy, with $k=\pi/d$ the lattice wavevector. Along $z$ is also present an harmonic potential with frequency  $\omega_z$=2$\pi\times$150~Hz.

To tune the interparticle interaction we vary the 1D scattering length $a_{1D}=a^{2}_{\perp}(1 - 1.03a/a_{\perp})/2a$, where $a_{\perp}=\sqrt{\hbar /m\omega_{\perp}}$ is fixed by the 2D lattice whereas the 3D scattering length $a$ can be adjusted at a Feshbach resonance \cite{Roati07} by using a magnetic field.
By varing the 1D scattering length, we tune both the Josephson energy $E_J$ and the Lieb-Liniger parameter $\gamma$. The Josephson energy is defined as $E_j=\hbar v_s/\sqrt{2}d$, where $v_s=\hbar\sqrt{\rho d^2/a_{1D}}/m^*$ is the sound velocity, $m^*=1.05~m$ is the effective mass in the lattice and $\rho$ is the density. The Lieb-Liniger is defined as $\gamma = 1 /(\rho_0 a_{1D})$, with $\rho_0$ being the tube peak density. In the limit of no lattice potential ($V=0$) and for small interactions ($\gamma\leq 10$) $\gamma $ can be related to the Luttinger parameter as $K\approx \pi/\sqrt{\gamma -\gamma^{3/2}/(2 \pi)}$ \cite{Nagerl}. In our inhomogeneous system both $E_J$ and $\gamma$ are calculated averaging over all tubes \cite{Boeris2016}, resulting in $E_J\approx$ 0.3 $\gamma^{1/4}$.
Also the mean filling $n$ changes with the interaction and it is found to scale as $n\approx \gamma^{-1/4}$. In our range of $\gamma$, when moving from weak to strong interactions, $n$ changes from about 2 to 1.

In the range of parameters we have studied, our system is always in the underdamped regime, $G<\omega_z$, indicating that phase slips are generated on timescales longer than the oscillation period. In this regime it is meaningful to study the dependence of $G$ on the maximum velocity $v$ reached during the first oscillation. $v$ corresponds to a trap displacement $\Delta z$ that can be suitably changed by varying a magnetic field gradient which partially compansates for gravity.

The critical velocity $v_c$ is measured according to the technique introduced in Ref.~\cite{Tanzi}. As the interaction is increased, we find that $v_c$ decreases as expected, varying from 7.1(3) to 5.4(6)~mm/s \cite{note1}.

From the momentum width at $t$=0 and the mean atom number per site, we are able to estimate the temperature $T$ via $k_BT=\hbar n \delta p_0/0.64 m^*d$ \cite{Gerbier,Gerbier04}, where the effective mass takes into account the presence of the shallow lattice. Since in our measurements $T$ is below the 1D degeneracy temperature $T_c\simeq50$~nK, the system is in the quasicondensate regime \cite{Petrov}.

All the error bars and uncertainties reported in the manuscript are statistical errors. A 30\% systematic uncertainty on the calibration of the atom number leads to additional systematic errors of 12\% on $\gamma$, 6\% on $E_j$, 12\% on $T$ and 6\% on $T/E_j$.\\


\begin{thebibliography}{99}
\bibitem{note2}  Another derivation of the relation between $\Gamma$ and $G$ is the following: The deceleration at the first maximum in the oscillations is $dv/dt$=$-Gv$. In terms of individual phase slips this can be written as $\delta v/\delta t$, where $\delta v$=$-h/mL$ is the deceleration following a phase slip of $2\pi$ in a chain of length $L$ and $\delta t^{-1}$=$\Gamma$.
\bibitem{note5} By fitting the data with power-low functions we measure exponents of $G$ ranging between 0.8 and 2, for the various interaction strengths explored in the experiment.
\end{thebibliography}
\end{document}